\documentclass[twoside,leqno,twocolumn]{article}  
\usepackage{ltexpprt} 
\usepackage{amsmath, amssymb}
\usepackage{enumerate}
\usepackage[square,sort,comma,numbers]{natbib}
\usepackage{subcaption}
\makeatletter
\renewcommand*{\@opargbegintheorem}[3]{\trivlist
          \item[\hskip \labelsep{\bfseries #1\ #2}] \textbf{(#3)}\ \itshape}
                  \makeatother

\newtheorem{assumption}{Assumption}

\newcommand{\tr}{TrustRank}
\usepackage{graphicx}

\begin{document}

\title{\Large Algebraic reputation model RepRank and its application to spambot detection \thanks{This work was supported by Russian Science Foundation Grant 14-11-00659}}
\author{G.V. Ovchinnikov \thanks{Skolkovo Institute of Science and Technology,
Skolkovo 143025, Russia and Institute of Design Problems in Microelectronics, Zelenograd Sovetskaya 3,  Moscow, 124365 (e-mail: ovgeorge@yandex.ru)} \\
  D.A. Kolesnikov \thanks{Skolkovo Institute of Science and Technology,
  Skolkovo 143025, Russia (e-mail: d4kolesnikov@yandex.ru)} \\
  \and
 I.V. Oseledets \thanks{Skolkovo Institute of Science and Technology,
 Skolkovo 143025, Russia and Institute of Numerical Mathematics, Russian Academy of Sciences,
ul. Gubkina 8, Moscow, 119333  Russia. (e-mail: ivan.oseledets@gmail.com)}
}
\date{}

\maketitle

%\pagenumbering{arabic}
%\setcounter{page}{1}%Leave this line commented out.

\begin{abstract} \small\baselineskip=9pt Due to popularity surge social networks became lucrative targets for
spammers and guerilla marketers, who are trying to game ranking systems and broadcast their messages at little to none cost.  Ranking systems, for example Twitter's Trends, can be gamed by scripted users also called bots, who are automatically or semi-automatically twitting essentially the same message.
Judging by the prices and abundance of supply from PR firms this is an easy to implement and widely used tactic, at least in Russian blogosphere. 
Aggregative analysis of social networks should at best mark those messages as spam or at least
correctly downplay their importance as they represent opinions only of a few, if dedicated, users. Hence bot detection plays a crucial role in social network mining and analysis.

In this paper we propose technique called RepRank which could be viewed as Markov chain based model for reputation propagation on graphs utilizing simultaneous trust and anti-trust propagation and provide effective numerical approach for its computation. 

Comparison with another models such as \tr$\quad$and some of its modifications on sample of 320000 Russian speaking Twitter users is presented. The dataset is presented as well.

\end{abstract}

\section{Introduction}

While concepts of 'good' and 'bad' are quite complex and may be considered subjective, applied to spam filtration problems they are
mapped to 'signal' and 'noise' correspondingly and became objective enough to build algorithms upon.

The following assumption, proposed in \cite{TR} was successfully used 
in graph based antispam algorithms:

\begin{assumption}[Approximate isolation of the good set]
\newline
Good graph vertices rarely link to the bad ones. 
\label{ass1}
\end{assumption}

This assumption leads to trust propagation scheme proposed in \cite{TR}: pages linked from good ones are almost good, 
the pages linked from those are slightly worse and so on. 
There is a similar approach to mistrust propagation \cite{ATR}: pages linking to bad pages are almost bad, the pages linking to those pages may be slightly better and so on.
The main difference being trust propagates forward (by graph edges direction) and mistrust propagates backward. 

\tr\, and other similar models for signal propagation on graphs can be viewed in random walker framework with random walkers carrying signal. 
In case of \tr\, it is trust charge equal to current vertex trust value. 
For some vertices marked by external verification process (also called oracle function) walkers with a priory set probability $\alpha$ take charge equal to $1$ and with probability $1-\alpha$ take charge equal to current charge of the vertex. 
Stationary distribution of such process with initial distribution
vector $d$ with $d_i=1$ if $i$-th vertex marked by the oracle as a good one and $d_i = 0$ otherwise is gives us \tr\, scores for whole graph:
\begin{equation}
        t = \alpha F t + (1 - \alpha) d,
\end{equation}
here $F$ is forward transition matrix which is column normalized adjacency matrix.

The logical continuation of \tr and anti-\tr is the combination of both models.  
Major advantage of combined trust and mistrust propagation approach 
is the ability to use 
both positive and negative signals. 

\cite{TDR} uses this approach by penalizing trust and distrust propagation form from not trustworthy and trustworthy vertices correspondingly. \cite{GBR} is analogous to \cite{TDR}, but stated in probabilistic framework. 

In this paper we further pursue the idea of simultaneous trust and mistrust propagation. In contrast with above-mentioned papers we combine trust and mistrust providing unified reputation score, called RepRank.

\section{Proposed model}
To each graph vertex we attach a trust score witch is negative for bad vertices and positive for a good ones.

Denote by $t_{+}$ and $t_{-}$ vectors obtained by zeroing negative and positive components in vector $t$ correspondingly.
Then, we search trust distribution $t$ to satisfy
\begin{equation}\label{reprank:eq1}
        t = \alpha_1 F t_{+} + \alpha_2 B t_{-} + \alpha_3 d,
\end{equation}
where $B$ is backward transition matrix which is
column normalized transposed adjacency matrix.
The solution of (\ref{reprank:eq1}) we will call RepRank. 
It exists, unique and continuously dependent on the initial distribution $d$ (see the Theorem \ref{theorem} for more details).
While usefulness of existence and uniqueness of the solution to the equation (\ref{reprank:eq1}) solution is beyond doubt we want to point out, that continuous dependence on oracle provided labeling $d$ is very nice as well. 
It protects RepRank from sudden "everything you knew is wrong" changes caused be addition of small portion of new data, mistakes and typos (a manual labeling is very tedious and error prone process). 
This follows natural intuition that once one have an idea of everyone's reputation a small changes should only correct, not shake the foundations of his worldview. 

\begin{theorem}
\label{theorem}
Let $F, B$ be $n \times n$ stochastic matrices from
(\ref{reprank:eq1}), $0 < \alpha_1, \alpha_2, \alpha_3 < 1$,
$\mathcal{P}_{+}$ be an operator which replaces all negative
component of the $n \times 1$ vector
by zeroes and $\mathcal{P}_{-}$ be operator which replaces all positive components of the
$n \times 1$ vector by zeroes.
Then mapping  $\mathcal{R}$ an $n \times 1$ vector $d$ to the $n
\times 1$ vector $t$ that the solution of the equation
\begin{equation}\label{reprank:eq2}
        t = \alpha_1 F   \mathcal{P}_{+}(t) + \alpha_2 B \mathcal{P}_{-}(t) + \alpha_3 d,
\end{equation}
has the next properties:
\begin{enumerate}
\item $R(d)$ exists for any vector $d$ from $\mathbf{R}^n$.
\item $R(d)$ is bijection mapping $\mathbf{R}^n$ to $\mathbf{R}^n$.
\item $R(d$) is Lipschitz continuous mapping.
\end{enumerate}
\end{theorem}
Proof:\\
One can use the following iterative process to find $t$:
\begin{equation*}\label{reprank:iter}
\begin{split}
    &  t^{(k+1)} = I(t^{(k)}),\\
    &I(t) = \alpha_1 F   \mathcal{P}_{+}(t) + \alpha_2 B  \mathcal{P}_{-}(t) + \alpha_3 d,
\end{split}
\end{equation*} 
with any initial vector $t_0$.
The mapping $I(t)$
is contractive on the metric space ($\mathbf{R}^n, \Vert \cdot \Vert_1$):
\begin{equation*}
\begin{split}
\Vert I(t_1) - I(t_2) \Vert_1 &=\\ 
= \Vert \alpha_1 F
(\mathcal{P}_{+}(t_1) - \mathcal{P}_{+}(t_2)) \\+\alpha_2 B
(\mathcal{P}_{-}(t_1) - \mathcal{P}_{-}(t_2)) \Vert_1 \leq \\ 
\leq \alpha_1 \Vert F \Vert_1 \Vert
\mathcal{P}_{+}(t_1) - \mathcal{P}_{+}(t_2)\Vert_1 + \\
\alpha_2 \Vert B \Vert_1
\Vert \mathcal{P}_{-}(t_1) - \mathcal{P}_{-}(t_2) \Vert_1 \leq \\
\leq \alpha_1 \Vert
\mathcal{P}_{+}(t_1) - \mathcal{P}_{+}(t_2)\Vert_1 + \\
\alpha_2 
\Vert \mathcal{P}_{-}(t_1) - \mathcal{P}_{-}(t_2) \Vert_1 \leq \\
\leq \max(\alpha_1 , \alpha_2) \Vert t_1 - t_2\Vert_1&,
\end{split}
\end{equation*}
where we used that $\Vert F\Vert_1 = \Vert B \Vert_1 = 1$ and
\begin{equation*}\label{1norm:eq}
\Vert t_1 - t_2\Vert_1 = \Vert\mathcal{P}_{+}(t_1) -
\mathcal{P}_{+}(t_2)\Vert_1 + \Vert \mathcal{P}_{-}(t_1) -
\mathcal{P}_{-}(t_2)\Vert_1.
\end{equation*}

So $I(t)$ is contractive mapping with coefficient less or equal to
$\max(\alpha_1 , \alpha_2) < 1$ and Banach fixed-point theorem
guarantees the existence and uniqueness of fixed point for it. Let
us denote that
fixed point by $R(d)$ and notice that it is the solution for equation
(\ref{reprank:eq1}).

Also $R(d)$ is Lipschitz continuous mapping with coefficient 
$$\frac{\alpha_3}{1- \max(\alpha_1,
   \alpha_2)}.$$
It can be proven in the following way:
\begin{equation*}
\begin{split}
 R(d_1) - R(d_2) &= \\
=\alpha_1 F (\mathcal{P}_{+}(R(d_1)) - \mathcal{P}_{+}(R(d_2)))& + \\
+\alpha_2 B (\mathcal{P}_{-}(R(d_1)) - \mathcal{P}_{-}(R(d_2))) &+\alpha_3 (d_1 - d_2) \\
\Vert R(d_1) - R(d_2) \Vert_1 &\leq \\ 
\leq \alpha_1 \Vert F \Vert_1 \Vert (\mathcal{P}_{+}(R(d_1)) -
\mathcal{P}_{+}(R(d_2))) \Vert_1& +\\
+\alpha_2 \Vert B \Vert_1 \Vert (\mathcal{P}_{-}(R(d_1)) -
\mathcal{P}_{-}(R(d_2))) \Vert_1 + \\ +\alpha_3 \Vert d_1 - d_2 \Vert d_2 \leq \\
\leq \max(\alpha_1, \alpha_2) \Vert R(d_1) - R(d_2) \Vert_1 &+ \alpha_3 \Vert d_1 - d_2 \Vert_1, 
\end{split}
\end{equation*}
therefore
\begin{equation*}
\Vert R(d_1) - R(d_2) \Vert_1 \leq \frac{\alpha_3}{1- \max(\alpha_1,
   \alpha_2)} \Vert d_1 - d_2 \Vert_1
.
\end{equation*}
The mapping $R(d)$ is injection because equality $R(d_1) = R(d_2)$ causes $d_1 =
d_2$ in equation (\ref{reprank:eq1}).
It is also a surjection because for any $t \in \mathbf{R}^n$ exists $d_t$ such that
$$d_t = \frac{1}{\alpha_3} (t - \alpha_1 F   \mathcal{P}_{+}(t) -
\alpha_2 B  \mathcal{P}_{-}(t)),$$ 
that equality $R(d_t) = t$ holds. 
\section{Experimental evaluation}

For our experiments we 
recursively crawled twitter using as seeds Russian speaking users we found in Twitter's Streaming API.
For each user we downloaded all people he follows, 'friends' in Twitter terminology. 
This friends graph has 326130 vertices and 2713369 nodes. We manually labeled 3124 users as spammers or a good ones. 
\footnote{The dataset can be obtained from \newline https://bitbucket.org/ovchinnikov/rutwitterdataset}

\begin{table}
  \centering
\begin{tabular}{ | l | c |}
  \hline
  Algorithm & Accuracy \\ 
  \hline
  \hline
  RepRank & 0.8833  \\
  \hline
  TrustRank & 0.851  \\
  \hline
  anti-TrustRank & 0.8636  \\
  \hline
\end{tabular}
\caption{Experimental results}
\label{tb:cmp}
\end{table}

We did a cross-validation with random subsampling splitting data set in two halves to test our algorithm against other single-score trust propagation algorithms, 
namely TrustRank and anti-TrustRank. The parameters for each algorithm were chosen to maximize accuracy. 

The results provided in the Table \ref{tb:cmp}. As can be seen our
algorithm outperforms both \tr$\quad$ and anti-\tr.

\begin{figure}[t]
\centering
\begin{subfigure}{0.49\linewidth} \centering
    \includegraphics[width=0.9\textwidth]{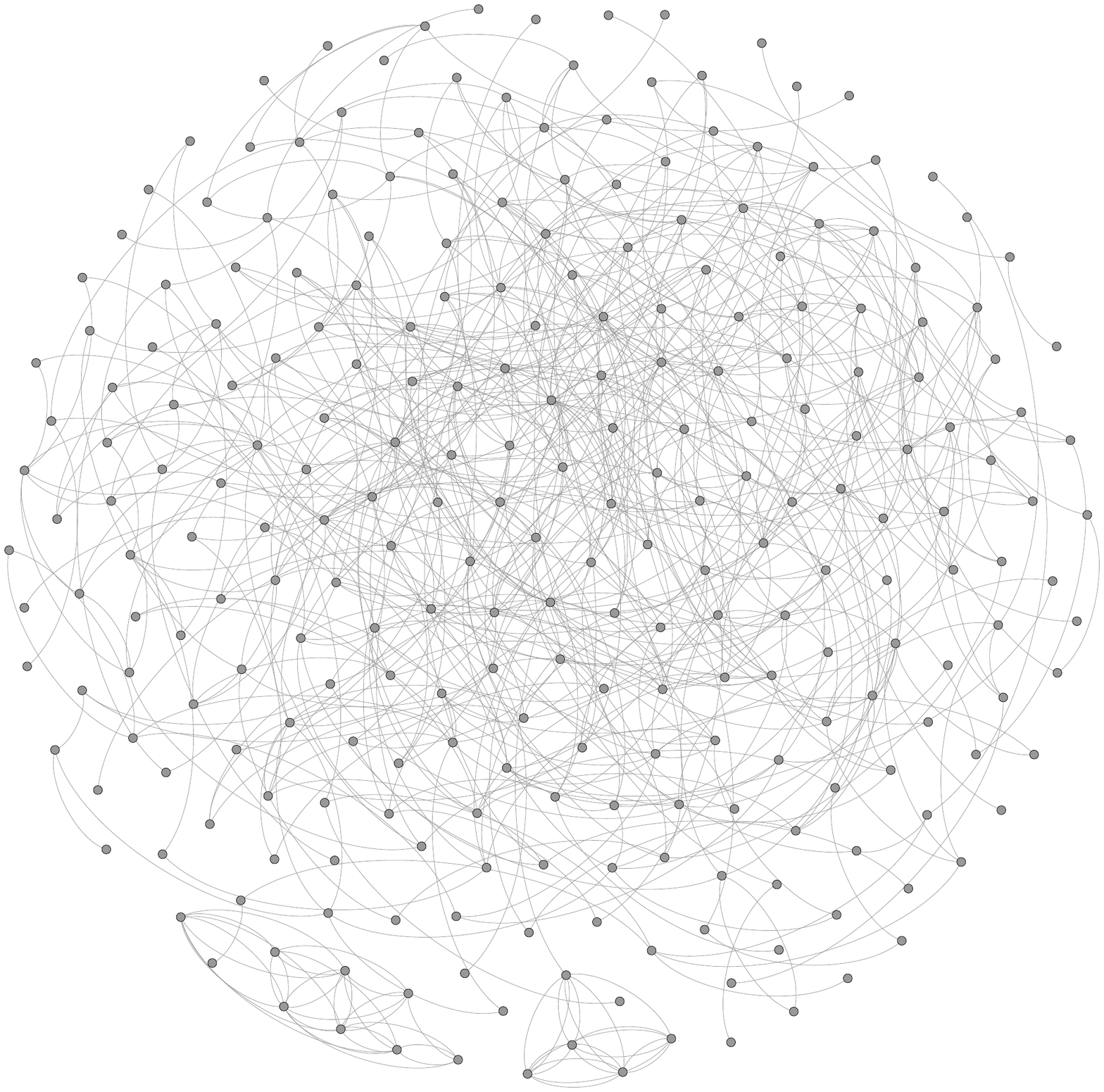}
\end{subfigure}
\begin{subfigure}{0.49\linewidth}
  \centering
    \includegraphics[width=0.9\textwidth]{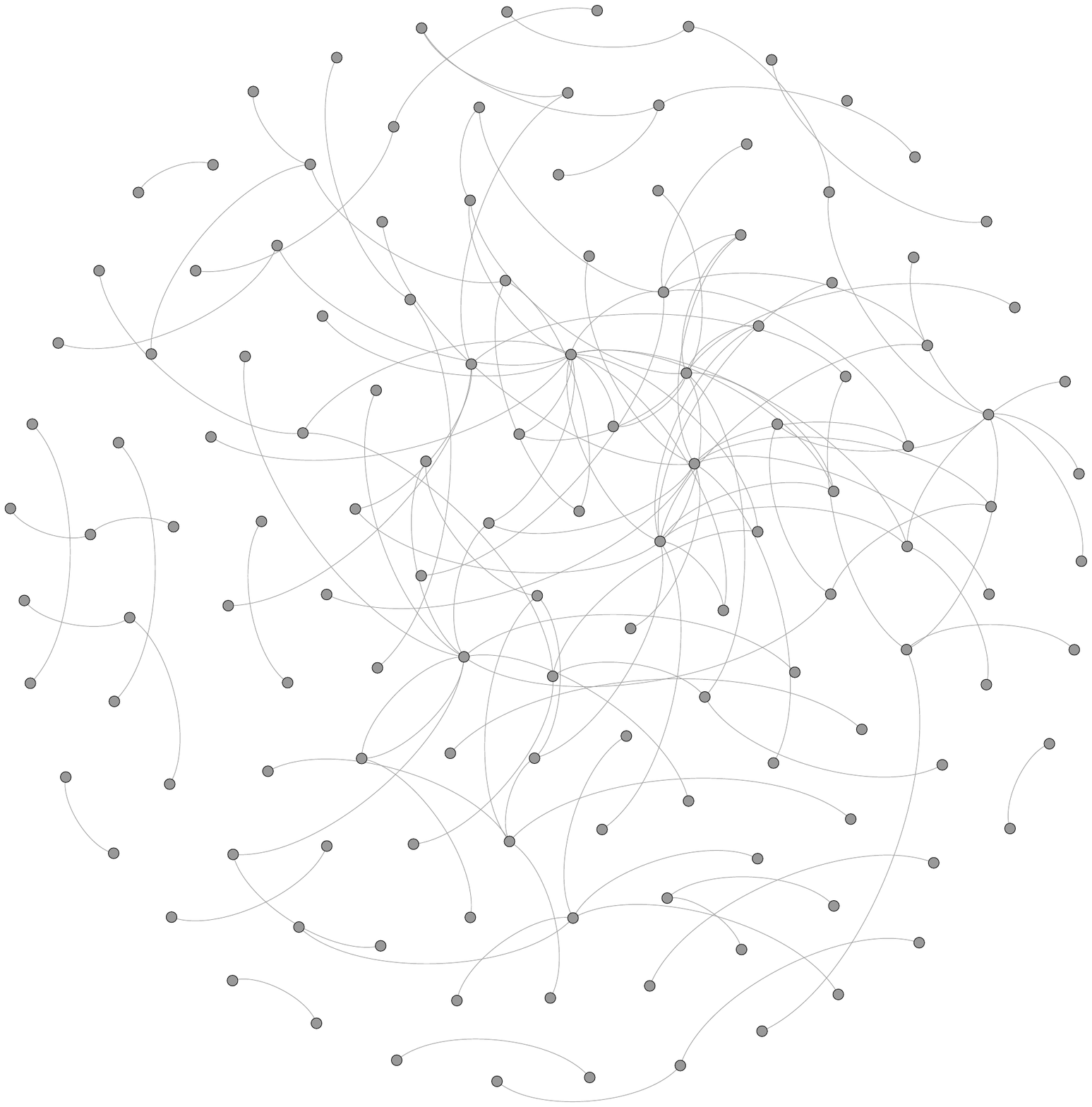}
\end{subfigure}
\caption{Connections between 300 vertices with highest in-degree (left) and 300 most reputable vertexes (right). Vertices not having connections within this group are omitted.  }
\label{fig:top300}
\end{figure}

\begin{figure}
    \centering
    \includegraphics[width=0.49\textwidth]{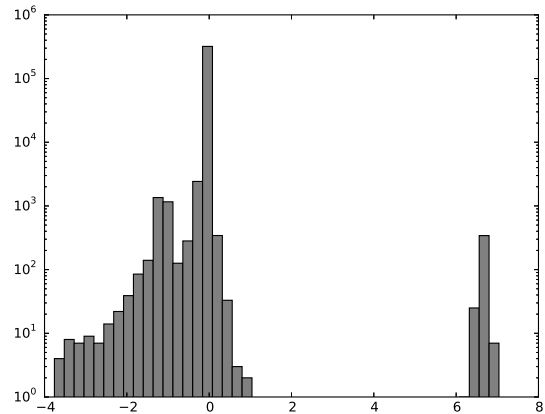}
    \caption{Distribution of reputation. RepRank score on the x-axis and logarithm of number of accounts on the y-axis. The gap is due to separation between masses and a handful of celebrities who are following each other. }
    \label{fig:rephist}
\end{figure}

It is interesting to note, that Russian-speaking part of Twitter is dominated by bots and spammers. 
According to our algorithm, out of 326130 accounts only 59691 (around 18\%) are managed by humans. Among 
those only 375 (around 0.1\%) correspond to high-reputation, profiles of prominent public figures, government 
officials and organizations, reputable press and so on (see reputation distribution on Figure \ref{fig:rephist}).
Spammers are more active at following each other than a normal people (see Figure \ref{fig:top300}). 
A serious anti-spam effort is required if one wants to make use of Twitter's data, otherwise he will analyze noise, or worse, some botmaster's political views.

\section{Conclusion and further work}

The contribution of this paper is twofold. First, we proposed new reputation propagation algorithm which allows to use both bad and good vertices in its starting set and outperforms analogues. Second, we gathered a sample Russian Twitter social graph with manual labeling of good and bad seeds. 

Different normalization strategies along with regularizations (pagerank-style teleportations) could be used to further improve performance of proposed method. 

\bibliographystyle{plain}
\bibliography{references}

\end{document}